\documentclass[12pt]{iopart}

\usepackage{iopams}
\usepackage{graphicx}

\begin{document}

\title{Analytical eigenstates for the quantum Rabi model}

\author{Honghua Zhong$^{1,2}$, Qiongtao Xie$^{2,3}$, Murray T. Batchelor$^{4,5}$ and Chaohong Lee$^{2}$}

\address{$^{1}$ Department of Physics, Jishou University, Jishou 416000, China}

\address{$^{2}$ State Key Laboratory of Optoelectronic Materials and Technologies, School of Physics and Engineering, Sun Yat-Sen University, Guangzhou 510275, China}

\address{$^{3}$ School of Physics and Electronic Engineering, Hainan Normal University, Haikou 571158, China}

\address{$^{4}$ Centre for Modern Physics, Chongqing University, Chongqing 400044, China}

\address{$^{5}$ Mathematical Sciences Institute and Department of Theoretical Physics, Research School of Physics and Engineering, Australian National University, Canberra ACT 0200, Australia}

\ead{chleecn@gmail.com}


\begin{abstract}
We develop a method to find analytical solutions for the eigenstates of the quantum Rabi model.
These include symmetric, anti-symmetric and asymmetric analytic solutions given
in terms of the confluent Heun functions. Both regular and exceptional solutions are given in a unified form.
In addition, the analytic conditions for determining the energy spectrum are obtained.
Our results show that conditions proposed by Braak [Phys. Rev. Lett. \textbf{107}, 100401 (2011)] are a type of sufficiency condition for determining the regular solutions.
The well-known Judd isolated exact solutions appear naturally as truncations of the confluent Heun functions.
\end{abstract}

\section{Introduction}

The quantum Rabi model describes the simplest coherent coupling  of
a two-level system to a single bosonic mode~\cite{Rabi}. This simple
model has been widely used in atomic physics, optical physics and
condensed matter physics~\cite{Scully,Wagner}. It has been experimentally tested in
different physical systems, such as cavity systems~\cite{Raimond},
trapped ions~\cite{Liebfried}, quantum dots~\cite{Englund},
superconducting circuits~\cite{Nakamura,Wallraff,Niemczyk,Diaz,Chiorescu,Fedorov}
and photonic superlattices~\cite{Longhi,Crespi}. As an example in
the case of superconducting circuits, artificial two-level systems
are emulated by superconducting devices, and single bosonic modes
are provided by LC oscillators or transmission line resonators~\cite{Blais}.
Due to the tunability of physical parameters, it is possible to test some important theoretical
predictions~\cite{Irisha,Irishb,Werlang,Ashhab,Hausinger,Casanova,Albert, Altintas} in experiments.

Recently, Braak obtained some analytical solutions for the quantum Rabi model and
derived the conditions for determining eigen-energies~\cite{Braak, Braakb}.
Braak's solutions and his conditions for determining eigen-energies have
stimulated extensive research interest on the quantum Rabi model and
similar models~\cite{Wolf,Wolfb,Hirokawa,Chen,Yu, Ziegler,Moroz,Maciejewski,Travenec,Gardas1,Gardas2, Zhang,Chilingaryan,Albert2012}.
Several different approaches have been proposed to solve this type of model,
such as Bogoliubov operators~\cite{Chen}, perturbation theory~\cite{Yu},
continued fractions~\cite{Ziegler,Moroz} and Wronskians~\cite{Maciejewski}.
In addition, various generalizations of the Rabi model, including a multi-photon Rabi model~\cite{Travenec, Gardas1},
two-mode Rabi model~\cite{Zhang}, multi-atom Rabi model~\cite{Chilingaryan} and an $N$-state Rabi model~\cite{Albert2012}
have been investigated.

Braak's analytical solutions for the quantum Rabi model are obtained
in the Bargmann space of analytical functions, where the system is
described by two coupled first-order ordinary differential
equations~\cite{Bargmann,Swain,kus,Reika,Reikb,Koc}. Braak showed
that the parity symmetry plays a key role in the integrability of
the Rabi model. He found an analytical solution and then derived the
conditions for determining the energy spectrum with the help of
parity symmetry. It has been shown that the energy spectrum of the
quantum Rabi model includes regular and exceptional parts.
The exceptional parts are related to the well-known Judd isolated
exact solutions which exist for some specific parameters~\cite{Judd}.
However, Braak's regular solutions and the corresponding conditions
are not applicable to Judd's exact solutions. Despite extensive
discussion of the quantum Rabi model, several fundamental problems
remain unresolved. Those that we address here are: (a) what are the complete analytical
solutions of the quantum Rabi model? and (b) is there a unified
approach which is applicable for both the regular and exceptional solutions?

In this paper we present a systematic investigation of the
analytical solutions for the quantum Rabi model and its
generalization. We suggest a different method to reduce the quantum Rabi model into operator-type differential equations and then obtain a set of analytical solutions including symmetric,
anti-symmetric and asymmetric solutions. These analytical solutions are constructed in terms of the Heun confluent function.
Three different types of conditions for the energy spectrum of the quantum Rabi model have been found.
The conditions due to the symmetric and anti-symmetric solutions are in principle applicable for both the regular and exceptional parts of the energy spectrum.
For the asymmetric solutions, the conditions are derived without the help of the underlying parity symmetry.
In particular, we show that Braak's conditions are a type of sufficiency condition for the energy spectrum of the Rabi model.

This paper is organized as follows. In section 2, we derive the coupled ordinary differential equations for the eigenvalue problem and give
two different types of analytical solutions in terms of the Heun confluent functions. In sections 3, 4 and 5, we present the symmetric,
anti-symmetric and asymmetric solutions and their corresponding eigenvalues. In section 6, we give Judd's isolated exact solutions in
the form of truncated confluent Heun functions. In the last section, we briefly summarize and discuss our results.

\section{Eigenvalue problem and its analytical solutions}

We consider the quantum Rabi model described by the Hamiltonian~\cite{Rabi},
\begin{equation}
H/\hbar =\omega a^{\dagger}a+\Delta
\sigma_z+g\sigma_x(a^{\dagger}+a).\label{ha}
\end{equation}
Here $a$ ($a^{\dagger}$) is the destruction (creation) operator for
a bosonic mode of frequency $\omega$, $\sigma_{x,z}$ are Pauli
matrices for the two-level system, $2\Delta$ is  the energy
difference between the two levels, and $g$ denotes the coupling
strength between the two-level system and the bosonic mode.
For simplicity and without loss of generality we set $\hbar=1$ and $\omega=1$.

The eigenstates $|\psi\rangle$ for the model can be written
\begin{equation}
|\psi\rangle = \psi_1(a^{\dagger})|0\rangle \left|\uparrow\right\rangle
+\psi_2(a^{\dagger})|0\rangle \left|\downarrow\right\rangle, \label{state}
\end{equation}
where $\psi_{1,2}$ are analytical functions of the creation operator
$a^{\dagger}$, $|0\rangle$ is the vacuum state for the bosonic mode,
and $\left|\uparrow\right\rangle$ and $\left|\downarrow\right\rangle$ are the eigenstates
of $\sigma_z$ with eigenvalues $1$ and $-1$.
The eigenvalue equation $H|\psi\rangle=E|\psi\rangle$ gives
\begin{equation}
([H,\psi_1]+\psi_1H-E\psi_1)|0\rangle \left|\uparrow\right\rangle
+([H,\psi_2]+\psi_2H-E\psi_2)|0\rangle \left|\downarrow\right\rangle=0.
\end{equation}
Using the relations $a|0\rangle=0$, $[a^{\dagger},\psi_{1,2}]=0$
and $[a,\psi_{1,2}]=\frac{d\psi_{1,2}}{da^{\dagger}}$,
the operator functions $\psi_{1,2}$ are seen to satisfy the operator-type
differential equations~\cite{Wu1,Wu2,zhong}
\begin{eqnarray}
z\frac{{d\psi_1}}{{d z}}+g\left(\frac{{d\psi_2}}{{d z}}+z\psi_2\right)+\Delta \psi_1 =E\psi_1, \label{diffa} \\
z\frac{{d\psi_2}}{{d z}}+g\left(\frac{{d\psi_1}}{{d z}}+z\psi_1\right)-\Delta
\psi_2 =E\psi_2,\label{diffb}
\end{eqnarray}
with $z=a^{\dagger}$. Since the operators in
equations (\ref{diffa}) and (\ref{diffb}) mutually commute,
they can formally be regarded as complex numbers. If we
make use of the linear combinations
$f_1=\psi_1+\psi_2$ and $f_2=\psi_1-\psi_2$,  the Schr\"{o}dinger
equation $H|\psi\rangle=E|\psi\rangle$ reduces to two coupled
first-order differential equations for $f_1(z)$ and $f_2(z)$~\cite{Wu1,Wu2},
\begin{eqnarray}
\frac{{df_1}}{{dz}}  &=& \frac{E-g z}{z+g}f_1- \frac{\Delta}{z+g}f_2,\label{ceqa}\\
\frac{{df_2}}{{dz}}  &=& \frac{E+g z}{z-g}f_2-
\frac{\Delta}{z-g}f_1.\label{ceqb}
\end{eqnarray}
These equations are equivalent to the ones in the Bargmann space of analytical functions~\cite{Bargmann,Swain,kus,Reika,Reikb,Koc}.

Through eliminating $f_2$, the coupled first-order differential equations~(\ref{ceqa}) and (\ref{ceqb})
are equivalent to a second-order differential equation for $f_1(z)$,
\begin{equation}
\frac{{d^2f_1}}{{dz^2}}+p(z)\frac{{df_1}}{{dz}}+q(z)f_1=0,\label{seqa}
\end{equation}
with
\begin{eqnarray}
p(z) &=& \frac{(1-2E-2g^2)z-g}{z^2-g^2}, \\
q(z) &=&\frac{-g^2 z^2+g z+E^2 -g^2-\Delta^2}{z^2-g^2}.
\end{eqnarray}
To derive the analytical solutions for equation (\ref{seqa}), we write the solutions in the form
\begin{eqnarray}
&\textrm{Type-I:}~~~& f_1(z)=\e^{-gz} \phi_1(x_1),\\
&\textrm{Type-II:}~~& f_1(z)=\e^{gz} \phi_2(x_2).
\end{eqnarray}
In the following, we derive the analytical solutions for
$\phi_1(x_1)$ and $\phi_2(x_2)$.

\subsection{Type-I solutions}

By introducing the new variable $x_1=(g-z)/2g$ and substituting $f_1(z)=\e^{-gz} \phi_1(x_1)$ into equation (\ref{seqa}),
we find that $\phi_1(x_1)$ obeys a confluent Heun equation~\cite{Ronveaux, Slavyanov} \footnote{Interestingly,
this kind of confluent Heun equation also appears in the semi-classical form of the quantum Rabi
model~\cite{Xie,Fillion}.}
\begin{equation}
\frac{d^2\phi_1}{dx_1^2} +\left(\alpha_1+\frac{\beta_1+1}{x_1}+\frac{\gamma_1+1}{x_1-1}\right)\frac{d\phi_1}{dx_1}
+\frac{\mu_1 x_1+\nu_1}{x_1(x_1-1)}\phi_1 =0, \label{heun1}
\end{equation}
with $\mu_1=\delta_1+\alpha_1(\beta_1+\gamma_1+2)/2$ and
$\nu_1=\eta_1+\beta_1/2+(\gamma_1-\alpha_1)(\beta_1+1)/2$.
The parameters $\alpha_1$, $\beta_1$, $\gamma_1$, $\delta_1$ are given
by $\alpha_1=4g^2$, $\beta_1=-(E+g^2+1)$, $\gamma_1=-(E+g^2)$,
$\delta_1=-2g^2$, and
$\eta_1=-3g^4/2+(1-2E)g^2/2+(E^2+E-2\Delta^2+1)/2$.

The confluent Heun equation has a local
Frobenius solution around $x_1=0$, which is known as the confluent
Heun function
$\varphi(x_1)=\textrm{HC}(\alpha_1,\beta_1,\gamma_1,\delta_1,\eta_1,x_1)
=\sum_{n=0}^\infty h_n x_1^n$, where the coefficients $h_n$ are
defined from the three-term recurrence relation  $A_n h_n=B_n
h_{n-1}+C_n h_{n-2}\;(n\geq 1)$ with the initial conditions $h_0=1$
and $h_{-1}=0$. Here $A_n=1+\beta_1/n$,
$B_n=1+(\beta_1+\gamma_1-\alpha_1-1)/n
+[\eta_1-\beta_1/2+(\gamma_1-\alpha_1)(\beta_1-1)/2]/n^2$, and
$C_n=[\delta_1+\alpha_1(\beta_1+\gamma_1)/2+\alpha_1(n-1)]/n^2$.
Another linearly independent solution is
$\widetilde{\varphi}_1(z)
=x_1^{-\beta_1}\textrm{HC}(\alpha_1,-\beta_1,\gamma_1,\delta_1,\eta_1,x_1)$.
Since  $-\beta_1=(E+g^2+1)$ is not always a non-negative integer, the
second solution $\widetilde{\varphi}_1(z)$ is a non-physical solution.
Therefore, the physical solution for $f_1(z)$ is given as
\begin{equation}
f_1(z)=C_1 \, \e^{-gz} \, \textrm{HC}\left(\alpha_1,\beta_1,\gamma_1,\delta_1,\eta_1,\frac{g-z}{2g}\right),\label{f1a}
\end{equation}
where $C_1$ is a constant to be determined.

Similarly, one can eliminate $f_1$ and obtain a second-order
differential equation for $f_2(z)$. By using the transformation
$f_2(z)=\e^{-gz} \, \varphi_2(x_1)$, it is easy to find that
$\varphi_2(x_1)$ satisfies a confluent Heun equation with parameters
$\alpha_2=4g^2$, $\beta_2=-(E +g^2)$, $\gamma_2=-(E+g^2+1)$,
$\delta_2=2g^2$ and
$\eta_2=-3g^4/2-(3+2E)g^2/2+(E^2+E-2\Delta^2+1)/2$. Correspondingly,
the analytical solution for $f_2(z)$ is
\begin{equation}
f_2(z)=C_2 \, \e^{-gz} \, \textrm{HC}\left(\alpha_2,\beta_2,\gamma_2,\delta_2,\eta_2,\frac{g-z}{2g}\right),\label{f2a}
\end{equation}
where $C_2$ is also a constant to be determined. Substitution of
$f_1(z)$ and $f_2(z)$ into equation (\ref{ceqb}) gives
$C_2/C_1=\Delta/(E+g^2 )$. Alternatively, we may obtain $f_2(z)$ by
substituting equation (\ref{f1a}) into equation (\ref{ceqa}). Note that our
analytical solutions (\ref{f1a}) and (\ref{f2a}) are different
from the solutions given by Braak \cite{Braak}.

\subsection{Type-II solutions}

Similar to the above working for the Type-I solutions, we
introduce the new variable $x_2=(g+z)/2g$ and substitute
$f_1(z)= \e^{gz} \, \phi_2(x_2)$ into equation (\ref{seqa}) to find that
$\phi_2(x_2)$ obeys the confluent Heun equation
\begin{equation}
\frac{d^2\phi_2}{dx_2^2} +\left(\alpha_3+\frac{\beta_3+1}{x_2}+\frac{\gamma_3+1}{x_2-1}\right)\frac{d\phi_2}{dx_2}
+\frac{\mu_3 x_2+\nu_3}{x_2(x_2-1)}\phi_2=0, \label{heun2}
\end{equation}
where $\mu_3=\delta_3+\alpha_3(\beta_3+\gamma_3+3)/2$ and
$\nu_3=\eta_3+\beta_3/2+(\gamma_3-\alpha_3)(\beta_3+1)/2$. The various
parameters are given by $\alpha_3=\alpha_1=4g^2$,
$\beta_3=\gamma_1=-(E+g^2)$, $\gamma_3=\beta_1=-(E +g^2+1)$,
$\delta_3=-\delta_1=2g^2$ and
$\eta_3=\eta_1+\delta_1=-3g^4/2-(3+2E)g^2/2+(E^2+E -2\Delta^2+1)/2$.

We thus obtain the analytical solutions for $f_1(z)$ and $f_2(z)$, which we write in the form
\begin{eqnarray}
\widetilde{f}_1(z)&=&D_1 \, \e^{gz} \, \textrm{HC}\left(\alpha_3,\beta_3,\gamma_3,\delta_3,\eta_3,\frac{g+z}{2g}\right),\label{f1b}\\
\widetilde{f}_2(z)&=&D_2 \, \e^{gz} \, \textrm{HC}\left(\alpha_4,\beta_4,\gamma_4,\delta_4,\eta_4,\frac{g+z}{2g}\right),
\label{f2b}
\end{eqnarray}
where $\alpha_4=\alpha_2=4g^2$, $\beta_4=\gamma_2=-(E+g^2)$,
$\gamma_4=\beta_2=-(E +g^2+1)$, $\delta_4=-\delta_2=-2g^2$, and
$\eta_4=\eta_2+\delta_2=-3g^4/2+(1-2E)g^2/2+(E^2+E-2\Delta^2+1)/2$.
Here $D_1/D_2=\Delta/(E+g^2)$ is obtained by substituting
$\widetilde{f}_1(z)$ and $\widetilde{f}_2(z)$ into equation (\ref{ceqa}).
Obviously, the parameters satisfy $\alpha_{3,4}=\alpha_{2,1}$,
$\beta_{3,4}=\beta_{2,1}$, $\gamma_{3,4}=\gamma_{2,1}$,
$\delta_{3,4}=\delta_{2,1}$ and $\eta_{3,4}=\eta_{2,1}$.

The coupled first-order ordinary differential equations~(\ref{ceqa})
and (\ref{ceqb}) for the eigenvalue problem have an important
symmetry. Under the transformation of $z\rightarrow-z$, we have
$f_1(-z)=cf_2(z)$ and $f_2(-z)=cf_1(z)$ for a constant $c$ with
equations (\ref{ceqa}) and (\ref{ceqb}) remaining unchanged. Under these
conditions it is easy to see that $c=\pm 1$. In
the following sections, we shall present a complete set of
analytical solutions including the symmetric, anti-symmetric and
asymmetric solutions.

\section{Symmetric solutions and their corresponding eigenvalues}

In this section, we show how to construct symmetric analytical solutions for equations (\ref{ceqa}) and (\ref{ceqb}).
Using equations (\ref{f1a})-(\ref{f2b}), the symmetric solutions can be written
\begin{eqnarray}
f_1^{+}(z)&=& \e^{-gz} \, \textrm{HC}\left(\alpha_1,\beta_1,\gamma_1,\delta_1,\eta_1,\frac{g-z}{2g}\right)
\nonumber\\
&&+\frac{\Delta}{E+g^2} \, \e^{gz} \, \textrm{HC}\left(\alpha_2,\beta_2,\gamma_2,\delta_2,\eta_2,\frac{g+z}{2g}\right),
\label{f1c}\\
f_2^{+}(z)&=&\frac{\Delta}{E+g^2} \, \e^{-gz} \,
\textrm{HC}\left(\alpha_2,\beta_2,\gamma_2,\delta_2,\eta_2,\frac{g-z}{2g}\right)\nonumber\\
&&+ \e^{gz} \, \textrm{HC}\left(\alpha_1,\beta_1,\gamma_1,\delta_1,\eta_1,\frac{g+z}{2g}\right),
\label{f2c}
\end{eqnarray}
which clearly satisfy the relation $f_1^{+}(z)= f_2^{+}(-z)$. The corresponding eigenstates are given as
\begin{eqnarray}
|\psi\rangle=\frac{f_1^{+}(z)+f_2^{+}(z)}{2} \left|0\right\rangle \left|\uparrow\right\rangle +\frac{f_1^{+}(z)-f_2^{+}(z)}{2} \left|0\right\rangle \left|\downarrow\right\rangle.
\end{eqnarray}
As above solutions (\ref{f1c}) and (\ref{f2c})
must satisfy equations (\ref{ceqa}) and (\ref{ceqb}), we have
\begin{eqnarray}
K^{+}(E, z)&:=& \e^{-gz} \, G_1^{+}(E,z)- \Delta \, \e^{gz} \, G_2^{+}(E,z) \nonumber\\
&:=& \e^{-gz} \, G_3^{+}(E,z)+ \frac{\Delta}{E+g^2} \, \e^{gz} \, G_4^{+}(E,z) \nonumber\\
&=&0,\label{kplus}
\end{eqnarray}
where
\begin{eqnarray}
G_1^{+}(E, z)&=&F_1(E,z)+ \frac{\Delta}{E+g^2} \, \e^{2gz} \, F_4(E, z), \\
G_2^{+}(E,z)&=&F_3(E, z)+ \frac{\Delta}{E+g^2} \, \e^{-2gz} \, F_2(E,z),\\
G_3^{+}(E, z)&=&F_1(E,z)- \Delta \, \e^{2gz} \, F_3(E, z), \\
G_4^{+}(E,z)&=&F_4(E, z)- \Delta \, \e^{-2gz} \, F_2(E,z) ,
\end{eqnarray}
with
\begin{eqnarray}
F_1(E,z)&=&(E+g^2)\, \textrm{HC}\left(\alpha_1,\beta_1,\gamma_1,\delta_1,\eta_1,\frac{g-z}{2g}\right)\nonumber\\
&&+\frac{g+z}{2g} \textrm{HC}' \left(\alpha_1,\beta_2,\gamma_1,\delta_1,\eta_1,\frac{g-z}{2g}\right),\\
F_2(E,z) &=&\textrm{HC}\left(\alpha_2,\beta_2,\gamma_2,\delta_2,\eta_2,\frac{g-z}{2g}\right),\\
F_3(E,z)&=&\textrm{HC}\left(\alpha_1,\beta_1,\gamma_1,\delta_1,\eta_1,\frac{g+z}{2g}\right),\\
F_4(E,z)&=&(E-g^2-2g z)\textrm{HC}\left(\alpha_2,\beta_2,\gamma_2,\delta_2,\eta_2,\frac{g+z}{2g}\right)\nonumber\\
&&-\frac{g+z}{2g} \textrm{HC}'\left(\alpha_2,\beta_2,\gamma_2,\delta_2,\eta_2,\frac{g+z}{2g}\right).
\end{eqnarray}
Here $\textrm{HC}'(\alpha,\beta,\gamma,\delta,\eta,x)$
denotes the derivative of the confluent Heun function
$\textrm{HC}(\alpha,\beta,\gamma,\delta,\eta,x)$ with respect to $x$.

Following~\cite{Braak}, the condition $K^+(E, z)=0$ holds for arbitrary values of $z$ if and only if $E$ corresponds to an
eigenvalue in the energy spectrum of the quantum Rabi model \footnote{The analogous statement applies to the condition $K^-(E, z)=0$ in the next section.}.
It therefore follows that equation (\ref{kplus}) is the condition for determining the energy spectrum of the quantum Rabi model.
Mathematically, the Heun confluent function
$\textrm{HC}(\alpha,\beta,\gamma,\delta,\eta,x)$ appearing in these solutions is convergent within the circle $|x|<1$, and  this leads to $z$ satisfying $|z-g|<2g$ and $|z+g|<2g$.
Numerically, we can determine the energy spectrum by taking a larger set of different values of $z$ satisfying $|z-g|<2g$ and $|z+g|<2g$ and then finding the common intersection of sets of the roots of $K^{+}(E,z)=0$ as a function of $E$ with a given value of $z$. However, we observe that it is inconvenient to find the spectrum numerically according to this procedure.
The reason is that, for any given value of $z$, $K^{+}(E, z)$ is very small for a wide range of $E$. It is thus difficult to identify the roots due to numerical errors.

\begin{figure}[tb]
\begin{center}
\centering \includegraphics{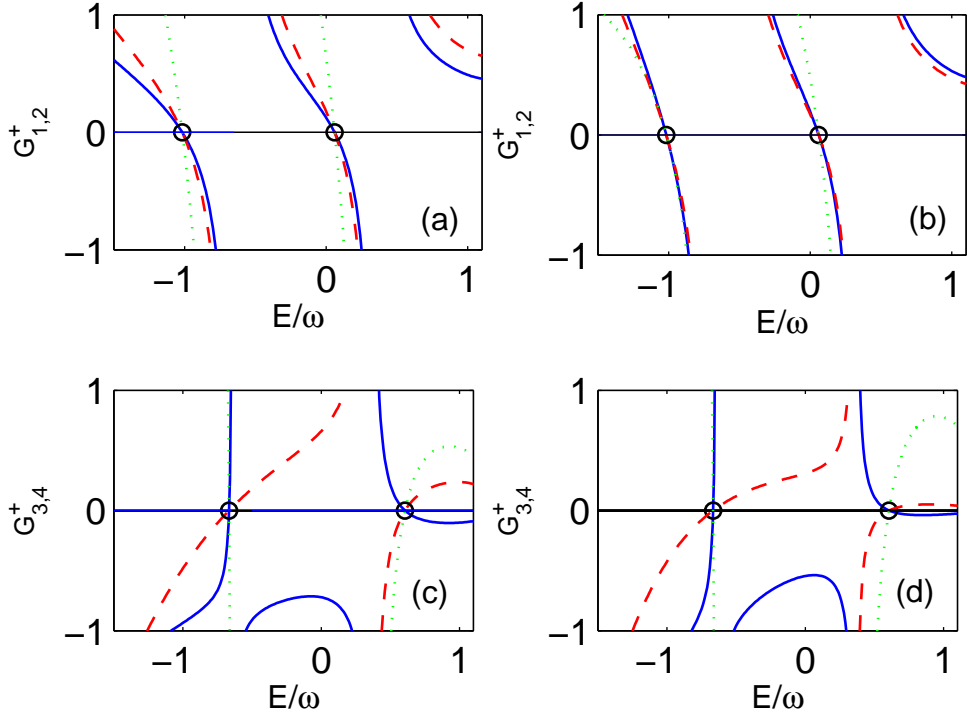}
\caption{Plots of $G_{1,2,3,4}^{+}(E, z)$ as functions of the energy $E/\omega$ for different values of $z$.
(a) and (b) $G_{1}^+(E,z)$ (solid  lines) and $G_{2}^+(E,z)$ (dashed  lines).
(c) and (d) $G_{3}^+(E,z)$ (solid lines) and $G_{4}^+(E,z)$ (dashed lines).
In (a) and (c) $z=0$ while in (b) and (d) $z=0.3$.
The other parameter values are $\omega=1$, $\Delta/\omega=0.7$ and $g/\omega=0.8$.
Here the zero axis lines define the conditions $G_{1,2,3,4}^{+}(E, z)=0$. The circles
correspond to the allowed energies. For comparison, we also
plot $G_-(E,z)$ and $G_+(E,z)$ (dotted lines) defined by Braak \cite{Braak}.
$G_-(E,z)$ is plotted in (a) and (b) and $G_+(E,z)$ in (c) and (d).
We see that $G_{1,2}^{+}(E, z)$ and $G_-(E,z)$ have the same roots.
Similarly for $G_{3,4}^{+}(E, z)$ and $G_+(E,z)$.}\label{fig1}
\end{center}
\end{figure}

However,  we can also obtain two sets of sufficient conditions from $K^{+}(E,z)=0$, namely the pair of weaker conditions
\begin{equation}
G_1^+(E,z)=G_2^+(E,z)=0 \quad \textrm{and} \quad G_3^+(E,z)=G_4^+(E,z)=0.
\end{equation}
It is clear that if these conditions are satisfied, then $K^{+}(E,z)=0$.
In figure~\ref{fig1}, we plot
$G_{1,2,3,4}^{+}(E, z)$ as functions of $E$ for two different values of $z$.
Our numerical results show that for a given value of $z$, $G_{1}^{+}(E,z)$ ($G_{3}^{+}(E, z)$) and $G_{2}^{+}(E, z)$
($G_{4}^{+}(E, z)$) have the same roots of $E$, which correspond to
the crossing points of the curves and the zero axis.

The common roots can be easily understood as follows. From $\Delta/(E+g^2) F_4(E,z)=\Delta F_3(E,z)$ and
$\Delta^2/(E+g^2)F_2(E,z)=F_1(E,z)$, we obtain $G_1^+(E,z)= \e^{2gz} \, \Delta G_2^+(E,z)$ and
$G_3^+(E,z)=-\e^{2gz} \, \Delta G_4^+(E,z)/(E+g^2)$, so that $G_{1}^{+}(E, z)$ ($G_{3}^{+}(E, z)$) and $G_{2}^{+}(E, z)$
($G_{4}^{+}(E, z)$) must have the same roots.

It is important to note that these roots correspond to parts of the allowed energies of the quantum Rabi model, marked by the circles.
It is clearly seen that the conditions $G_{1,2}^{+}(E, z)=0$ and $G_{3,4}^{+}(E, z)=0$ each only give part of the energy eigenvalue spectrum,
and together they give all of the energy eigenvalues given by Braak \cite{Braak}.
Here for comparison in figure~\ref{fig1}, we have also plotted the curves $G_-(E,z)$ and $G_+(E,z)$ defined by Braak.
It is found that $G_{1,2}^{+}(E, z)$ and $G_-(E,z)$ have the same roots.
$G_{3,4}^{+}(E, z)$ and $G_+(E,z)$ also share the same roots.
It is clearly seen that our results agree with Braak's results for the allowed energies.

\section{Anti-symmetric solutions and their corresponding eigenvalues}

In this section, we show how to construct anti-symmetric analytical solutions for equations (\ref{ceqa}) and (\ref{ceqb}).
By using equations (\ref{f1a})-(\ref{f2b}), the anti-symmetric solutions can be constructed in the form
\begin{eqnarray}
f_1^{-}(z)&=& \e^{-gz} \, \textrm{HC}\left(\alpha_1,\beta_1,\gamma_1,\delta_1, \eta_1,\frac{g-z}{2g}\right)
\nonumber\\
&&-\frac{\Delta}{E+g^2} \, \e^{gz}  \, \textrm{HC}\left(\alpha_2,\beta_2,\gamma_2,\delta_2, \eta_2,\frac{g+z}{2g}\right),\label{f1d}\nonumber\\
\\
f_2^{-}(z)&=&\frac{\Delta}{E+g^2} \, \e^{-gz} \, \textrm{HC}\left(\alpha_2,\beta_2,\gamma_2,\delta_2, \eta_2,\frac{g-z}{2g}\right)\nonumber\\
&&- \e^{gz} \, \textrm{HC}\left(\alpha_1,\beta_1,\gamma_1,\delta_1, \eta_1,\frac{g+z}{2g}\right),
\label{f2d}
\end{eqnarray}
which satisfy $f_1^{-}(z)=-f_2^{-}(-z)$. The corresponding eigenstates are given as
\begin{eqnarray}
|\psi\rangle=\frac{f_1^{-}(z)+f_2^{-}(z)}{2} \left|0\right\rangle \left|\uparrow\right\rangle +\frac{f_1^{-}(z)-f_2^{-}(z)}{2} \left|0\right\rangle \left|\downarrow\right\rangle.
\end{eqnarray}
Substituting above solutions (\ref{f1d}) and (\ref{f2d}) into equations (\ref{ceqa}) and (\ref{ceqb}), we obtain
\begin{eqnarray}
K^{-}(E, z)&:=& \e^{-gz} \, G_1^{-}(E,z)+ \Delta \, \e^{gz} \,G_2^{-}(E,z) \nonumber\\
&:=& \e^{-gz} \, G_3^{-}(E,z)- \frac{\Delta}{E+g^2} \, \e^{gz} \,G_4^{-}(E,z) \nonumber\\
&=&0,
\end{eqnarray}
where
\begin{eqnarray}
G_1^{-}(E, z)&=&F_1(E,z)- \frac{\Delta}{E+g^2}  \, \e^{2gz} \, F_4(E, z), \\
G_2^{-}(E,z)&=&F_3(E, z)- \frac{\Delta}{E+g^2} \, \e^{-2gz} \, F_2(E,z),\\
G_3^{-}(E, z)&=&F_1(E,z)+ \Delta  \, \e^{2gz} \, F_3(E, z), \\
G_4^{-}(E,z)&=&F_4(E, z)+\Delta \, \e^{-2gz} \, F_2(E,z).
\end{eqnarray}
For this case the two sets of sufficient conditions for $K^{-}(E,z)=0$ are
\begin{equation}
G_1^-(E,z)=G_2^-(E,z)=0 \quad \textrm{and} \quad G_3^-(E,z)=G_4^-(E,z)=0.
\end{equation}

\begin{figure}[tb]
\begin{center}
\includegraphics[width=\columnwidth]{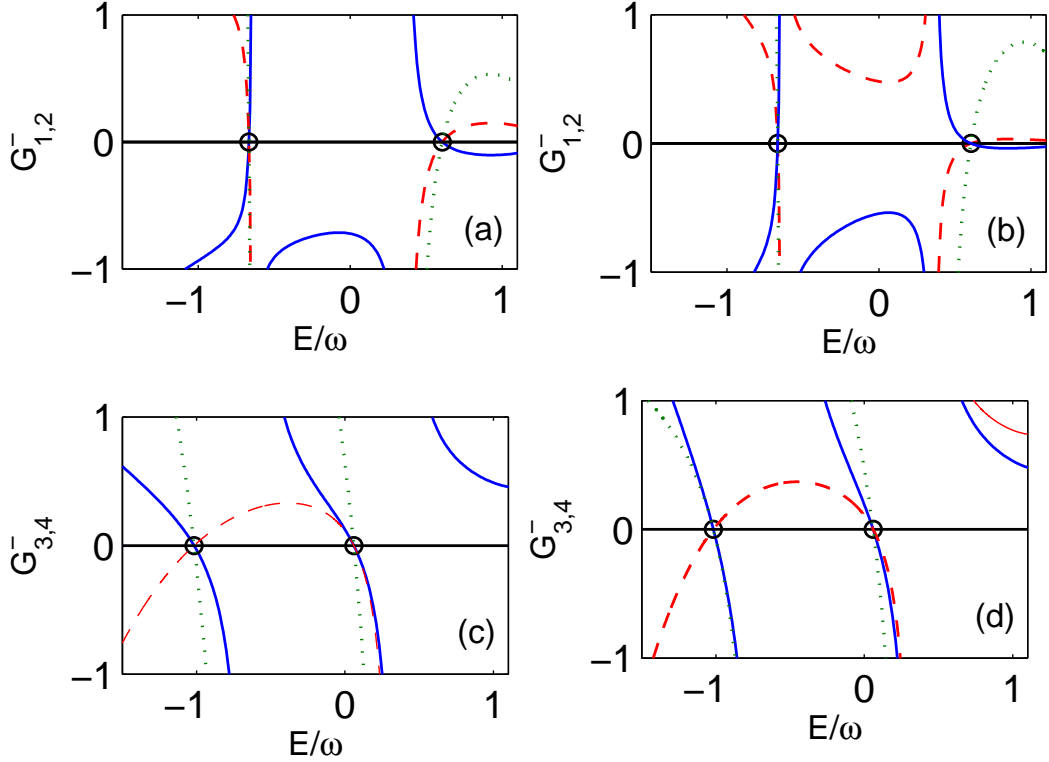}
\caption{Plots of $G_{1,2,3,4}^{-}(E, z)$ as functions of the energy $E/\omega$ for different values of $z$.
(a) and (b) $G_{1}^-(E,z)$ (solid  lines) and $G_{2}^-(E,z)$ (dashed  lines).
(c) and (d) $G_{3}^-(E,z)$ (solid lines) and $G_{4}^-(E,z)$ (dashed lines).
In (a) and (c) $z=0$ while in (b) and (d) $z=0.3$.
The other parameter values are $\omega=1$, $\Delta/\omega=0.7$ and $g/\omega=0.8$.
Here the zero axis lines define the conditions $G_{1,2,3,4}^{-}(E, z)=0$. The circles
correspond to the allowed energies. For comparison, we also
plot $G_-(E,z)$ and $G_+(E,z)$ (dotted lines) defined by Braak \cite{Braak}.
$G_-(E,z)$ is plotted in (a) and (b) and $G_+(E,z)$ in (c) and (d).
We see that $G_{1,2}^{-}(E, z)$ and $G_-(E,z)$ have the same roots.
Similarly for $G_{3,4}^{-}(E, z)$ and $G_+(E,z)$.}\label{fig2}
\end{center}
\end{figure}

We plot $G_{1,2,3,4}^{-}(E, z)$ as a function of $E$ for two different values of $z$ in figure~\ref{fig2}.
Here we can use $G_1^-(E,z)=-\e^{2gz} \, \Delta G_2^-(E,z)$ and
$G_3^-(E,z)=\e^{2gz} \,\Delta G_4^-(E,z)/(E+g^2)$ to show that
$G_{1}^{-}(E, z)$ and $G_{2}^{-}(E, z)$ have the same roots of $E$ for a given value of $z$.
And similarly for $G_{3}^{-}(E, z)$ and $G_{4}^{-}(E, z)$.
We find that the conditions $G_{1,2}^{-}(E, z)=0$ and $G_{3,4}^{-}(E, z)=0$ each give part of the
energy eigenvalue spectrum and together they give all of the energy eigenvalues.
Also shown in figure~\ref{fig2} are the curves $G_-(E,z)$ and $G_+(E,z)$ given by Braak \cite{Braak}.
Here $G_{3,4}^{-}(E, z)$ give the same energy eigenvalues as $G_+(E,z)$,
while $G_{1,2}^{-}(E, z)$ give the same results as $G_-(E,z)$.

To conclude this section, we note that the functions $G_{3,4}^{+}(E, z)$ and $G_{3,4}^{-}(E, z)$ can be
regarded as defining a set of conditions for the energy eigenvalues. Indeed, the
conditions $G_3^{\pm}(E,z)=0$ and $G_4^{\pm}(E,z)=0$ can be obtained using our analytic solutions and
following the procedure used by Braak \cite{Braak}. We first take
$f_1(z)= \e^{-gz} \, \textrm{HC}(\alpha_1,\beta_1,\gamma_1,\delta_1,\eta_1,\frac{g-z}{2g})$
or
$\widetilde{f}_1(z)= \e^{g z} \, \textrm{HC}(\alpha_1,\gamma_1,\beta_1,-\delta_1,\eta_1+\delta_1,\frac{g+z}{2g})$
and then substitute into equation (\ref{ceqa}) to obtain
$\Delta f_2(z)= \e^{-gz} \, F_1(E,z)$ or
$\Delta \widetilde{f}_2(z)= \e^{g z} \, F_4(E,z)$.
The conditions $G_3^{\pm}(E, z)=0$ or $G_4^{\pm}(E,z)=0$
follow from the relations $f_2(z)=\pm f_1(-z)=\pm \, \e^{g z} \, F_3(E, z)$ or
$\widetilde{f}_2(z)=\pm \widetilde{f}_1(-z)=\pm \, \e^{-g z} \, F_2(E, z)$.
Our results show that Braak's conditions are a type of sufficiency condition for the
regular parts of the energy spectrum of the quantum Rabi model.

\section{Asymmetric solutions and their corresponding eigenvalues}

In addition to the above symmetric and anti-symmetric solutions, we may construct the following two sets of asymmetric analytical solutions
\begin{eqnarray}
f_1^1(z)&=& \e^{-gz} \, \textrm{HC}\left(\alpha_1,\beta_1,\gamma_1,\delta_1,\eta_1,\frac{g-z}{2g}\right),\label{f1e}\\
f_2^1(z)&=&\frac{\Delta}{E+g^2}\, \e^{-gz} \, \textrm{HC}\left(\alpha_2,\beta_2,\gamma_2,\delta_2,\eta_2,\frac{g-z}{2g}\right),
\label{f2e}
\end{eqnarray}
and
\begin{eqnarray}
f_1^2(z)&=&\frac{\Delta}{E+g^2} \, \e^{gz} \, \textrm{HC}\left(\alpha_2,\beta_2,\gamma_2,\delta_2, \eta_2,\frac{g+z}{2g}\right),
\label{f1f}\\
f_2^2(z)&=& \e^{gz} \, \textrm{HC}\left(\alpha_1,\beta_1,\gamma_1,\delta_1, \eta_1,\frac{g+z}{2g}\right),
\label{f2f}
\end{eqnarray}
for equations (\ref{ceqa}) and (\ref{ceqb}).
For these solutions, $f_1^{1,2}$ and $f_2^{1,2}$ themselves do not obey the parity symmetry, but they relate to each other by   $f_1^1(z)=f_2^2(-z)$ and $f_2^1(z)=f_1^2(-z)$.

\begin{figure}[htb]
\begin{center}
\includegraphics[width=\columnwidth]{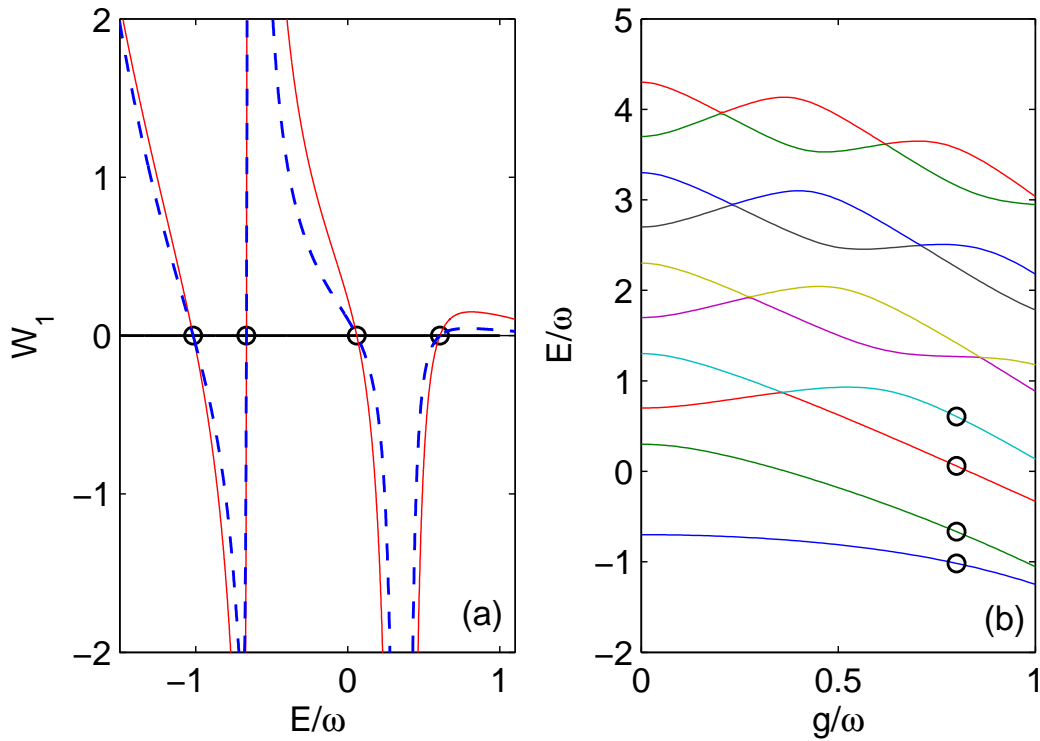}
\caption{(a) The Wronskian $ W_{1}(E,z)$ as a function of $E/\omega$
for $z=0$ (solid line) and $z=0.5$ (dashed lines) with $\omega=1$,
$\Delta/\omega=0.7$ and $g/\omega=0.8$. The circles represent the crossing points of
the curve $W_{1}(E,z)$ and the zero axis line, corresponding to the
allowed energies. (b) The spectrum of the quantum Rabi model for the same
parameter values as in (a). The circles correspond to the circles in (a).}\label{fig3}
\end{center}
\end{figure}

We now show how to determine the energy eigenvalues corresponding to asymmetric solutions. For the regular parts of the energy spectrum of the quantum Rabi model, $(f_1^{1}, f_2^1)$ and $(f_1^{2}, f_2^2)$ should represent the
same eigenstates
\begin{eqnarray}
|\psi\rangle=\frac{f_1^{1}(z)+f_2^{2}(z)}{2} \left|0\right\rangle \left|\uparrow\right\rangle +\frac{f_1^{1}(z)-f_2^{1}(z)}{2} \left|0\right\rangle \left|\downarrow\right\rangle.
\end{eqnarray}
This requires that the Wronskians for $f_1^{1,2}$
and $f_2^{1,2}$ must be zero, i.e.,
\begin{eqnarray}
W_1(E,z) &=& \frac{df_{1}^{1}}{dz}f_{1}^{2}-f_{1}^{1}\frac{df_{1}^{2}}{dz}=0, \\
W_2(E,z) &=& \frac{df_{2}^{1}}{dz}f_{2}^{2}-f_{2}^{1}\frac{df_{2}^{2}}{dz}=0.
\end{eqnarray}
Since $f_1^1(z)=f_2^2(-z)$ and $f_2^1(z)=f_1^2(-z)$ we have
$W_{1}(E,-z)=W_{2}(E,z)$. Therefore, we need only consider
$W_{1}(E,z)$. In figure~\ref{fig3}(a) we plot $W_{1}(E,z)$ as a
function of $E$ for two different values of $z$.
The crossing points of the curve $W_{1}(E,z)$ and the zero axis line correspond to the
allowed energy eigenvalues of $E$, marked by the circles. In
figure~\ref{fig3}(b), the lowest parts of the energy spectrum of the
quantum Rabi model are plotted. The circles in figure~\ref{fig3}(b)
correspond to the circles in figure~\ref{fig3}(a). In this situation,
the condition $W_{1}(E,z)=0$ gives all the regular parts of the
energy values. It is important to note that this condition for the
regular spectrum of the Rabi model  is obtained without the help of
the underlying symmetry of the model.

\section{Judd's isolated exact solutions}

It is well known that, for some particular values of $\Delta$ and $g$, the quantum Rabi model (\ref{ha}) supports exact analytical solutions known as Judd's isolated exact solutions~\cite{Judd}. In this section, we show that Judd's exact solutions and the conditions for determining their energy eigenvalues can be obtained easily from the asymmetric solutions.

The confluent Heun functions are holomorphic, i.e., polynomial functions with an infinite number of terms.
However, under the conditions
$\delta=-(N+(\gamma+\beta+2)/2)\alpha$ and $h_{N+1}=0$ with nonnegative integers $N$, the Heun functions may be terminated as truncated solutions in the form of a finite series, with
$\textrm{HC}(\alpha,\beta,\gamma,\delta,\eta,x)=\sum_{n=0}^Nh_nx^n$~\cite{Ronveaux,Slavyanov}.
Here the condition $\delta=-(N+(\gamma+\beta+2)/2)\alpha$ gives parts of the energy spectrum, while the condition $h_{N+1}=0$ imposes a special relation between the system parameters $\Delta$ and $g$. Below, we give two simple examples.

\subsection{Example 1}

To truncate the confluent Heun functions in equations (\ref{f1e}) and (\ref{f2e}) as finite series, we require
$E=N_1-g$ and $h_{N_1+1}=0$ for equation (\ref{f1e}) and $E=N_2+1-g^2$ and
$h_{N_2+1}=0$ for equation (\ref{f2e}). For the case of $N_2=0$ and
$N_1=1$, we have $\Delta^2+4g^2=1$ and $E=1- g^2$. The corresponding
solutions are
\begin{eqnarray}
f_1^{1,1}(z)&=& \e^{-gz} \, (1-2g^2+2g z),\\
f_2^{1,1}(z)&=&\Delta \, \e^{-gz}.
\end{eqnarray}
The corresponding eigenstate for the above truncated solution is
\begin{eqnarray}
\left|\psi\right\rangle&=& \e^{g^2/2} \, \left[(1+\Delta-2g^2)\left|-g\right\rangle +2g\sqrt{L_1(-g^2)} \left|-g,1\right\rangle\right]\left|\uparrow\right\rangle  \nonumber\\
&&+ \e^{g^2/2} \, \left[(1-\Delta-2g^2)\left|-g\right\rangle
+2g\sqrt{L_1(-g^2)} \left|-g,1\right\rangle\right]
\left|\downarrow\right\rangle,
\end{eqnarray}
where $\left|-g\right\rangle$ denotes the coherent state,
$\left|-g,m\right\rangle$ denotes the photon added coherent
state~\cite{Agarwal}, and $L_m(x)$ is the Laguerre polynomial of
order $m$. In general, $f_{1,2}^{1}(z)\left|0\right\rangle$ are a
superposition of the  photon added coherent state
$\left|-g,m\right\rangle$, and $f_{1,2}^{2}(z)\left|0\right\rangle$
are a superposition of the photon added coherent state
$\left|g,m\right\rangle$ with nonnegative integer $m$.

Similarly,
for $f_1^2(z)$ and $f_2^2(z)$ with $\Delta^2+4g^2=1$ and $E=1- g^2$,
equations (\ref{f1f}) and (\ref{f2f}) give
\begin{eqnarray}
f_1^{2,1}(z)&=&\Delta \, \e^{gz},\\
f_2^{2,1}(z)&=& \e^{gz} (1-2g^2-2g z).
\end{eqnarray}
The corresponding eigenstate is
\begin{eqnarray}
\left|\psi\right\rangle&=& \e^{g^2/2} \, \left[(1+\Delta-2g^2)\left|g\right\rangle -2g\sqrt{L_1(-g^2)} \left|g,1\right\rangle\right]\left|\uparrow\right\rangle  \nonumber\\
&&+ \e^{g^2/2} \, \left[(\Delta-1+2g^2)\left|g\right\rangle
+2g\sqrt{L_1(-g^2)} \left|g,1\right\rangle\right]
\left|\downarrow\right\rangle.
\end{eqnarray}

With the two degenerated asymmetric solutions for Judd's isolated states, $(f_1^{1,1}, f_2^{1,1})$ and $(f_1^{2,1}, f_2^{2,1})$, one can construct the symmetric solution
\begin{eqnarray}
f_1^{+,1}(z)&=& f_1^{1,1}(z)+f_1^{2,1}(z)= \e^{-gz} \, (1-2g^2+2g z)+\Delta \, \e^{gz},\\
f_2^{+,1}(z)&=& f_2^{1,1}(z)+f_2^{2,1}(z)= \Delta \, \e^{gz} + \e^{gz} \, (1-2g^2-2g z).
\end{eqnarray}
From this symmetric solution, we have
\begin{eqnarray}
G_1^+(E,z)&=&1-4g^2+\Delta \, \e^{2gz} \, (1-2g^2-2gz),\\
G_2^+(E,z)&=&(1-2g^2-2gz)+\Delta \, \e^{-2gz},\\
G_3^+(E,z)&=&1-4g^2-\Delta \, \e^{2 gz} \, (1-2g^2-2gz), \\
G_4^+(E,z)&=&(1-2g^2-2gz)-\Delta \, \e^{-2gz}.
\end{eqnarray}
It is clear that for any given value of $z$, we have $K^+(E,z)=0$, but $G_{1,2,3,4}^+(E,z)\neq 0$.
This implies that the weaker conditions $G_{1,2,3,4}^+(E,z)=0$ are only applicable for the regular parts of the energy spectrum of the quantum Rabi model.

In a similar way, one can also construct the finite anti-symmetric solution
\begin{eqnarray}
f_1^{-,1}(z)&=& f_1^{1,1}(z)-f_1^{2,1}(z)= \e^{-gz} \, (1-2g^2+2g z)-\Delta \, \e^{gz},\\
f_2^{-,1}(z)&=& f_2^{1,1}(z)-f_2^{2,1}(z)= \Delta \, \e^{-gz} \, - \e^{gz} \, (1-2g^2-2g z).
\end{eqnarray}
For this anti-symmetric solution, we have
\begin{eqnarray}
G_1^-(E,z)&=&1-4g^2-\Delta \, \e^{2gz} \, (1-2g^2-2gz), \\
G_2^-(E,z)&=&(1-2g^2-2gz)-\Delta \, \e^{-2gz},\\
G_3^-(E,z)&=&1-4g^2+\Delta \, \e^{2gz} \, (1-2g^2-2gz), \\
G_4^-(E,z)&=&(1-2g^2-2gz)+\Delta \, \e^{-2gz}.
\end{eqnarray}
Similarly, it is clear
that for any given value of $z$, we have  $K^-(E,z)=0$, but
$G_{1,2,3,4}^-(E,z)\neq 0$. Our analytical results show that the
conditions $K^{\pm }=0$ are in principle applicable for both the
regular and exceptional parts of the energy spectrum of the quantum
Rabi model. However, the weaker conditions $G_{1,2,3,4}^{\pm}(E,z)=0$ are only applicable
for the regular parts of the energy spectrum.

\subsection{Example 2}

As a further case, for $N_2=1$ and $N_1=2$ in  equations (\ref{f1e}) and
(\ref{f2e}), we have $32g^4+4(3\Delta^2-8)g^2+\Delta^4-5\Delta^2+4=0$ and $E=2- g^2$, with
the corresponding solutions
\begin{eqnarray}
f_1^{1,2}(z)&=& \e^{-gz} \, \left(1+c_1\frac{g-z}{2g}+c_2\frac{(g-z)^2}{4g^2}\right),\\
f_2^{1,2}(z)&=&\case12 {\Delta} \, \e^{-gz} \, \left(1+d_1\frac{g-z}{2g}\right),
\end{eqnarray}
where $c_1=\case12 \Delta^2-2$, $c_2=8g^4-8g^2+2g^2\Delta^2$ and
$d_1=4g^2+\Delta^2-4$. Similarly, from equations (\ref{f1f}) and (\ref{f2f}) we obtain
\begin{eqnarray}
f_1^{2,2}(z)&=&\case12 {\Delta} \, \e^{gz} \, \left(1+d_1\frac{g-z}{2g}\right),\\
f_2^{2,2}(z)&=& \e^{gz} \, \left(1+c_1\frac{g-z}{2g}+c_2\frac{(g-z)^2}{4g^2}\right).
\end{eqnarray}

For the above two examples of Judd isolated exact solutions, we always have $K^{\pm}(E,z)=0$ but $G^{\pm}_{1,2,3,4}\neq 0$ for any given value of $z$. From the analytical solutions, we see that  the energies
$E=1-g^2$ with $\Delta^2+4g^2=1$ and $E=2-g^2$ with
$32g^4+4(3\Delta^2-8)g^2+\Delta^4-5\Delta^2+4=0$ have two-fold
degeneracy, because the corresponding Wronskians are nonzero for any
given value of $z$ satisfying  $|z-g|<2g$ and $|z+g|<2g$. For
example, for $f_1^{1,1}=\e^{-gz} \, (1-2g^2+2gz)$ and
$f_1^{2,1}=\Delta \, \e^{gz} $, we have $W=df_1^{1,1}/dx
f_1^{2,1}-f_1^{1,1} df_1^{2,1}/dx=4\Delta g^2(g-z)\neq 0$.
Generally, Judd isolated exact solutions are associated
with level crossings between two energy levels with different symmetry~\cite{Judd}.

\section{Conclusion}

In conclusion, for the quantum Rabi model, we have developed a method to find the analytical solutions
for the eigenstates in terms of the Heun confluent functions. These include symmetric, anti-symmetric and asymmetric solutions.
An analytical method to derive the conditions for determining the energy spectrum has also been given.
The conditions for the symmetric and anti-symmetric solutions are in principle applicable for both the
regular and the exceptional parts of the energy spectrum. Two sets of sufficient conditions for the regular spectrum are obtained.
The conditions for the asymmetric solutions determining the regular parts of the energy spectrum have also been found.

We have shown that our general  results agree very well in comparison with Braak's results~\cite{Braak}.
Moreover, Braak's sufficiency conditions can be regarded as
sufficient conditions for obtaining the regular parts of the energy spectrum.
Our results also provide other forms of the conditions for the
regular parts of the energy spectrum.

For some specific values of the parameters, the solution in terms of Heun confluent functions can be terminated as a polynomial with a finite number of terms. These eigenstates correspond precisely to the Judd isolated exact solutions~\cite{Judd}.
Here we have shown that the regular and exceptional parts of the eigenspectrum are unified by the common description in terms of the Heun confluent functions. Our analytical method could  be extended to solve other generalised quantum Rabi models, such as the non-parity-symmetric quantum Rabi model~\cite{Braak,Gardas2}, the qubit-oscillator system~\cite{Forn}, and the asymmetric quantum Rabi model with different coupling strengths for the counter-rotating wave and rotating wave interactions~{\cite{Hioe,Shen}.

\ack
This work is supported by the NBRPC under Grant No. 2012CB821305, the NNSFC under Grants No. 11075223 and 11147021,  the Ph.D. Programs Foundation of Ministry of Education of China under Grant No. 20120171110022, the NCETPC under Grant No. NCET-10-0850, and the Hunan Provincial Natural Science Foundation under Grant No. 12JJ4010. M.T.B. is supported by the 1000 Talents Program of China. His work is also partially supported by the Australian Research Council.

\end{document}